# Search for the Jacobi Instability in Rapidly Rotating $^{46}$Ti* Nuclei *

A. Maj, M. Kmiecik, W. Królas, J. Styczeń

The Henryk Niewodniczański Institute of Nuclear Physics, Kraków, Poland

AND

A. Bracco, F. Camera, B. Million

Milano University and INFN, Milano, Italy

AND

J.J. Gaardhøje, B. Herskind

The Niels Bohr Institute, Copenhagen, Denmark

AND

M. Kicińska-Habior, J. Kownacki

Warsaw University, Warsaw, Poland

AND

W.E. Ormand

Lawrence Livermore National Laboratory, Livermore, USA

The possible existence of Jacobi shape transition in hot $^{46}$Ti at high angular momenta was investigated with the Giant Dipole Resonance exclusive experiments. The GDR spectra and the angular distributions are consistent with predictions of the thermal shape fluctuation model indicating elongated nuclear shapes.

PACS numbers: 21.10.Re; 24.30.Cz; 25.70.Gh

---

## 1. Introduction

Rapidly rotating nuclei may exhibit an abrupt change of equilibrium shape from an oblate non-collectively rotating ellipsoid to a triaxial or prolate rotating body. This phenomenon was suggested by Beringer and Knox [1] already 40 years ago, and formulated in semi-classical models by W. Świątecki and collaborators [2, 3, 4]. The suggestion was based on a finding done by C.G.J. Jacobi in 1834, that a stable equilibrium shape of gravitating mass rotating synchronously, changes abruptly at a certain angular momentum, from a slightly oblate spheroid to a triaxial ellipsoid rotating around its shortest axis. This phenomenon may be observed in light- and medium-mass nuclei, where the critical angular momentum for Jacobi transitions is well below the critical angular momentum for fission. For example in $^{46}$Ti, according to formulae in [4], the Jacobi transition should appear at the 29 $\hbar$, whereas the angular momentum at which the fission barrier vanishes is 40 $\hbar$. In the experiment, the Jacobi shapes of nuclei should manifest themselves in alteration of the observables in which the moment of inertia is involved. In the cold rapidly rotating nuclei, one should observe a giant backbend of the quadrupole transition energies [4, 5]. In hot nuclei, such very deformed shapes should be seen in the spectra of high-energy $\gamma$-rays from the decay of Giant Dipole Resonance (GDR), since the GDR strength function splits itself for deformed nuclei, and the splitting is proportional to the size of the quadrupole deformation. The signature for Jacobi shapes in $^{46}$Ti would be a pronounced high energy bump at E$_\gamma$=25 MeV in the strength function of the Giant Dipole Resonance (GDR) together with a rather large anisotropy of the $\gamma$-rays deexciting the GDR.

The first experimental indication for the Jacobi transition was obtained by the Seattle group [6] for $^{45}$Sc compound nuclei. In the GDR absorption cross-section from the inclusive experiment a "shoulder" at 25 MeV was found and attributed to a large effective prolate deformation. However, the simultaneously measured angular distributions did not show the expected behaviour.

## 2. Experimental results

To search for the Jacobi transition effects in exclusive experiments, we have studied the GDR decay from the hot $^{46}$Ti nucleus populated in the $^{18}$O+$^{28}$Si reaction at 98 MeV, using the accelerator facility of the Niels Bohr Institute at Risø (Denmark). The compound $^{46}$Ti nucleus, was produced with the excitation energy of 81 MeV, and an angular momentum distribution with l$_{max} \approx 32\hbar$. The high-energy $\gamma$-rays were detected in the HECTOR array [7] as a function of the $\gamma$-fold measured in the multiplicity filter HELENA [8]. Fig. 1 shows the experimental spectra for four regions of the γ-fold: 2-3, 4-5, 6-8 and 9-11. The upper row displays the measured spectra on the logarithmic scale, and the results of a statistical model fit. The calculations assume that the GDR strength function is of double Lorentzian type. To examine the details of the GDR strength function, the experimental spectra are converted into our best estimate of the photon absorption cross-section, and shown in a linear scale in the middle row of Fig. 1. This GDR cross-section is represented by the expression $F_{2L}*Y_{exp}/Y_{fit}$. In this expression, $Y_{exp}$ is the experimental spectrum and $Y_{fit}$ is the calculated spectrum assuming that the GDR decay can be represented by the double Lorentzian function $F_{2L}$; the latter being the best fit to the experimental spectrum. For the data gated by the highest folds, one can note a shoulder at $E_\gamma \approx 24$ MeV, similar to that seen in the $^{45}$Sc nucleus [6]. The quadrupole deformation parameter evaluated from the fitted centroids of the two GDR components is $\beta \approx 0.4$, consistent with the elongated Jacobi shapes. The $A_2$-coefficient spectra, as extracted from the angular distribution data, are displayed in the bottom row of Fig. 1. For the low folds (<9), the data are negative in both regions of the low and high energy GDR components, showing a similar peculiar behaviour observed for $^{45}$Sc [6]. In contrast, for the highest fold window (9-11), the behaviour of the $A_2$-coefficient changes and follows the expectations, i.e. being negative for the low energy component and positive for the high one. This can be explained, noting that the highest fold window corresponds to the relatively narrow angular momentum window of compound nucleus 29±4 $\hbar$, where the Jacobi transition is expected to happen. The average temperature sensed by the GDR in this angular momentum window is about 1.3 MeV. The lower folds, due to charged particle emission, correspond to almost the full angular momentum distribution of the compound nucleus [9], and are not sensitive to the angular momentum effects. It is interesting to notice a sharp peak around 12 MeV (observed also in [6]), being strong at low folds and vanishing with increase of the fold window. Most probably it is related to the $^{18}$O projectile break-up. Such events will appear at rather low $\gamma$-multiplicities and are filtered out at higher fold windows.

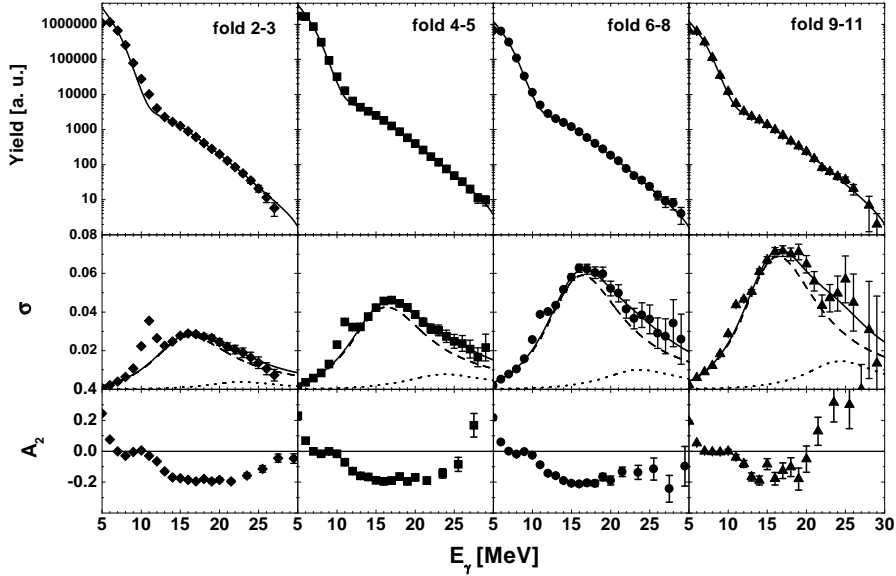

Fig. 1. Results of measurements of the GDR γ-decay in $^{46}$Ti*. Upper row: experimental spectra (on a logarithmic scale) for four regions of γ-fold (points) and results of the CASCADE code fits (lines). Middle row: the photo-absorption cross-section determined from the fits (solid lines) together with the two Lorentzian components (dashed lines); the photo-absorption cross-section was extracted from the measured spectra (points). Bottom row: experimental spectra of the $A_2$-coefficient of the angular distributions.

### 3. Interpretation

One can make simple estimates of the effective shape probed by the giant dipole oscillation by comparing the measured strength functions and $A_2$-spectra with the calculations at the fixed deformation parameter $\beta$=0.4, as obtained from the statistical model fit, and for 3 different non-axiality parameters $\gamma$: $0^o$ (prolate collective rotation), $-60^o$ (oblate non-collective rotation), and $-30^0$ (triaxial shape, rotating about shortest axis).

Those simple calculations, which may be interpreted as possible effective shapes, are displayed in Fig. 2 (left panels) in comparison with the experimental quantities, i.e. the GDR absorption cross-section and the $A_2$-values. As can be seen, the triaxial effective shape fits the best to the experimental data. One can then argue, that $^{46}$Ti nuclei have an effective deformation with $\beta \approx 0.4$ and the triaxial shape with $\gamma \approx -30^o$, corresponding to the ratio of major axes 1.6:1.25:1. Such shapes are expected for nuclei undergoing the Jacobi shape transition.

One knows, however, that nuclei do not possess a given stable shape at elevated temperatures, but rather have to be considered as a fluctuating shape ensemble [10, 11], described by the free energy contours F($\beta$, $\gamma$). A nucleus having the temperature T may posses a given shape parametrized by $\beta$ and $\gamma$ with a probability proportional to the the Boltzmann factor P($\beta$,$\gamma$) = exp(-F($\beta$, $\gamma$)/T). According to the thermal shape fluctuation model, each of these shapes contributes to the measured quantities with a weight given by the Boltzmann factor.

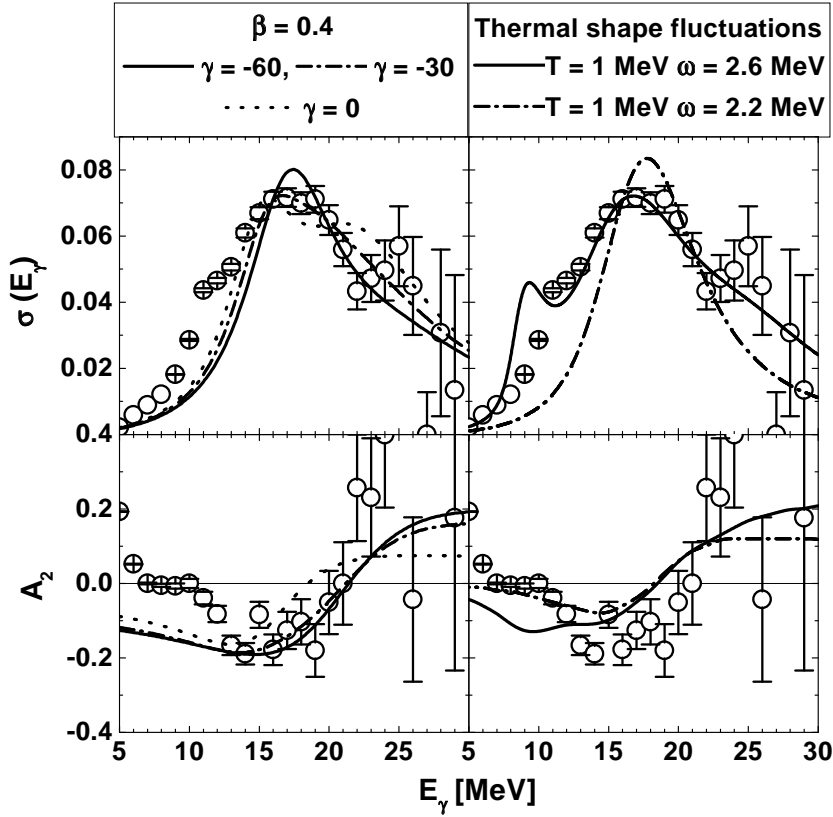

Fig. 2. Left panels: Experimental cross-section (top) and $A_2$-spectra (bottom) for the highest fold window together with simple calculations assuming the same deformation parameter $\beta$, but three different non-axiality parameters. Right panels: The experimental data compared to results from the thermal shape fluctuations model based on free energy calculations shown in Fig. 3.

The calculated contours of the free energy F(x,y) for $^{46}$Ti, where x=$\beta\cos\gamma$ and y=$\beta\sin\gamma$, are shown if Fig. 3. They were made using the formalism of the standard Nilsson-Strutinski procedure extended to finite temperatures [11, 12]. The calculations were performed for T=1 MeV for four values of the rotational frequency $\omega$=2.2, 2.4, 2.6 and 2.8 MeV, corresponding to I=23, 25, 28 and 30 $\hbar$, respectively. For rotational frequencies lower than $\omega$=2.5 MeV the free energy surface has an almost spherical minimum. For $\omega$=2.6 MeV ($\sim$28$\hbar$) two additional minima appear, one which is oblate with $\beta$=0.45, and the other, almost prolate, with $\beta$=1.2. The latter minimum disappears, when the rotational frequency increases. Therefore, this min-imum is probably related with the Jacobi shape. The Boltzmann factor distribution for the free energy at $\omega$=2.6 MeV was used for the thermal shape calculations (see e.g. [7]) of the GDR cross-section and $A_2$-spectra. The results are plotted as solid lines in the right panels of Fig. 2, and compared with the experimental results. One can see a good agreement for the cross-section, while the $A_2$-distribution would rather prefer more triaxial shapes. To demonstrate the influence of the Jacobi shape (the very deformed prolate minimum in the free energy landscape), we plot also the thermal shape calculations for $\omega$=2.2 MeV (dashed lines), where no Jacobi shapes are present. One can see that only when large deformations are involved, one can describe the experimental cross-section, thus supporting the indication of Jacobi transition present in hot and rapidly rotating $^{46}$Ti.

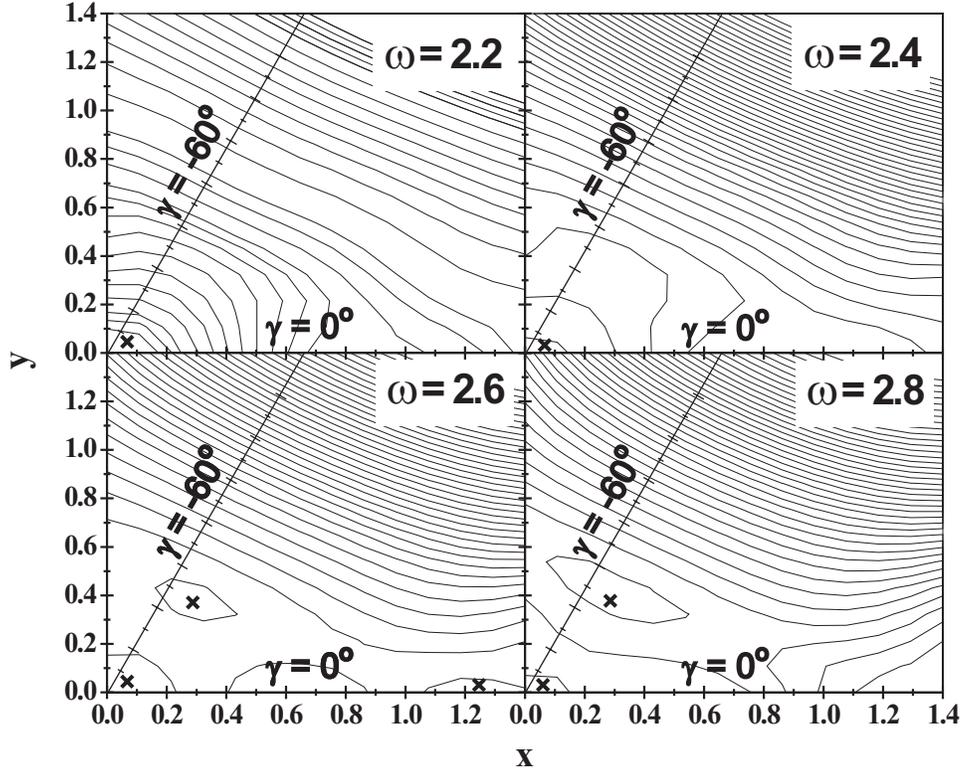

Fig. 3. Free energy F(x,y) calculations for $^{46}$Ti at temperature T=1 MeV and rotational frequencies $\omega$=2.2, 2.4, 2.6, 2.8 MeV (x=$\beta\cos\gamma$, y=$\beta\sin\gamma$). The main local minima are indicated with crosses.

## 4. Summary

The search for Jacobi shapes in rapidly rotating $^{46}$Ti$^*$ nuclei, based on high-energy $\gamma$-spectra from the decay of the GDR indicate, indeed, the presence of such a phenomenon. The measured quantities, such as the GDR cross-section and A$_2$-spectra, can be explained by the thermal shape fluctuation model based on free energy calculations in which the Jacobi transition appears at $\omega$=2.6 MeV. It would be worthwhile to extend this to the low energy region of discrete transitions, and look for the predicted giant backbend in $^{46}$Ti specifically.

This work was partially supported by the Polish State Committee for Scientific Research (KBN Grants No. 2 P03B 001 16 and 2 P03B 045 16), the Danish Science Foundation and the Italian Istituto Nazionale di Fisica Nucleare.